\documentclass[10pt,english,aps,amssymb,prl,twocolumn, showpacs]{revtex4-1}
\pdfoutput=1
\usepackage[utf8]{inputenc}
\usepackage{amsmath}
\usepackage{graphicx}
\usepackage{amssymb}
\usepackage{esint}
\usepackage{verbatim}
\usepackage{bm}
\usepackage{hyperref}

\makeatletter
\@ifundefined{textcolor}{}
{
 \definecolor{WHITE}{gray}{1}
 \definecolor{RED}{rgb}{1,0,0}
 \definecolor{GREEN}{rgb}{0,1,0}
 \definecolor{BLUE}{rgb}{0,0,1}
 \definecolor{CYAN}{cmyk}{1,0,0,0}
 \definecolor{MAGENTA}{cmyk}{0,1,0,0}
 \definecolor{YELLOW}{cmyk}{0,0,1,0}
 }

\usepackage{color}

\newcommand{\bra}[1]{\langle #1|}
\newcommand{\ket}[1]{|#1\rangle}

\newcommand{\brakets}[2]{\left\langle#1| #2 \right\rangle}
\renewcommand{\phi}{\varphi}
\renewcommand{\epsilon}{\varepsilon}

\renewcommand{\vec}[1]{{\bf #1}}

\DeclareMathOperator{\IM}{Im}
\DeclareMathOperator{\RE}{Re}
\DeclareMathOperator{\TR}{Tr}

\DeclareMathOperator{\sgn}{sgn}

\newcommand{\bs}{\boldsymbol}
\newcommand{\mc}{\mathcal}

\usepackage{babel}
\begin{document}
\title {Amorphous topological superconductivity in a Shiba glass}

\author{Kim P\"oyh\"onen}
\author{Isac Sahlberg}
\author{Alex Weststr\"om}
\author{Teemu Ojanen}
\affiliation{Department of Applied Physics (LTL), Aalto University, P.~O.~Box 15100,
FI-00076 AALTO, Finland }
\date{\today}
%
%
%
%
%
%


\maketitle
\bigskip{}

\textbf{Topological states of matter support quantized nondissipative responses and exotic quantum particles that cannot be accessed in common materials. The exceptional properties and application potential of topological materials have triggered a large-scale search for new realizations. Breaking away from the popular trend focusing almost exclusively on crystalline symmetries, we introduce the Shiba glass as a platform for amorphous topological quantum matter. This system consists of an ensemble of randomly distributed magnetic atoms on a superconducting surface. The collection of magnetic moments gives rise to subgap Yu-Shiba-Rusinov states that form a topological superconducting phase at critical density despite a complete absence of spatial order.  Experimental signatures of the amorphous topological state can be obtained  by STM measurements probing the topological edge mode. Our discovery demonstrates the physical feasibility of amorphous topological quantum matter and presents a concrete route to fabricating new topological systems from  nontopological materials with random dopants.}

Topological states are characterized by integer-valued invariants \cite{volovik, bernevig} that remain robust in the presence of imperfections. While topological properties can be studied independently of local order, spatial symmetries play a central role in virtually all material realizations. This is emphasized by the fact that the theoretical search for new topological materials extensively employs band structures and reciprocal space. While topological states are generically robust to disorder which breaks spatial symmetries, this is typically established by treating the disorder as additional feature in a well-defined clean system. Even topological Anderson insulators \cite{li, groth}, where moderate disorder actually gives rise to nontrivial topological properties, rely crucially on a specific band structure of the clean system. The concept of disorder, almost by definition, implies the existence of an underlying ordered reference state. 

\begin{figure}
\includegraphics[width=0.9\columnwidth]{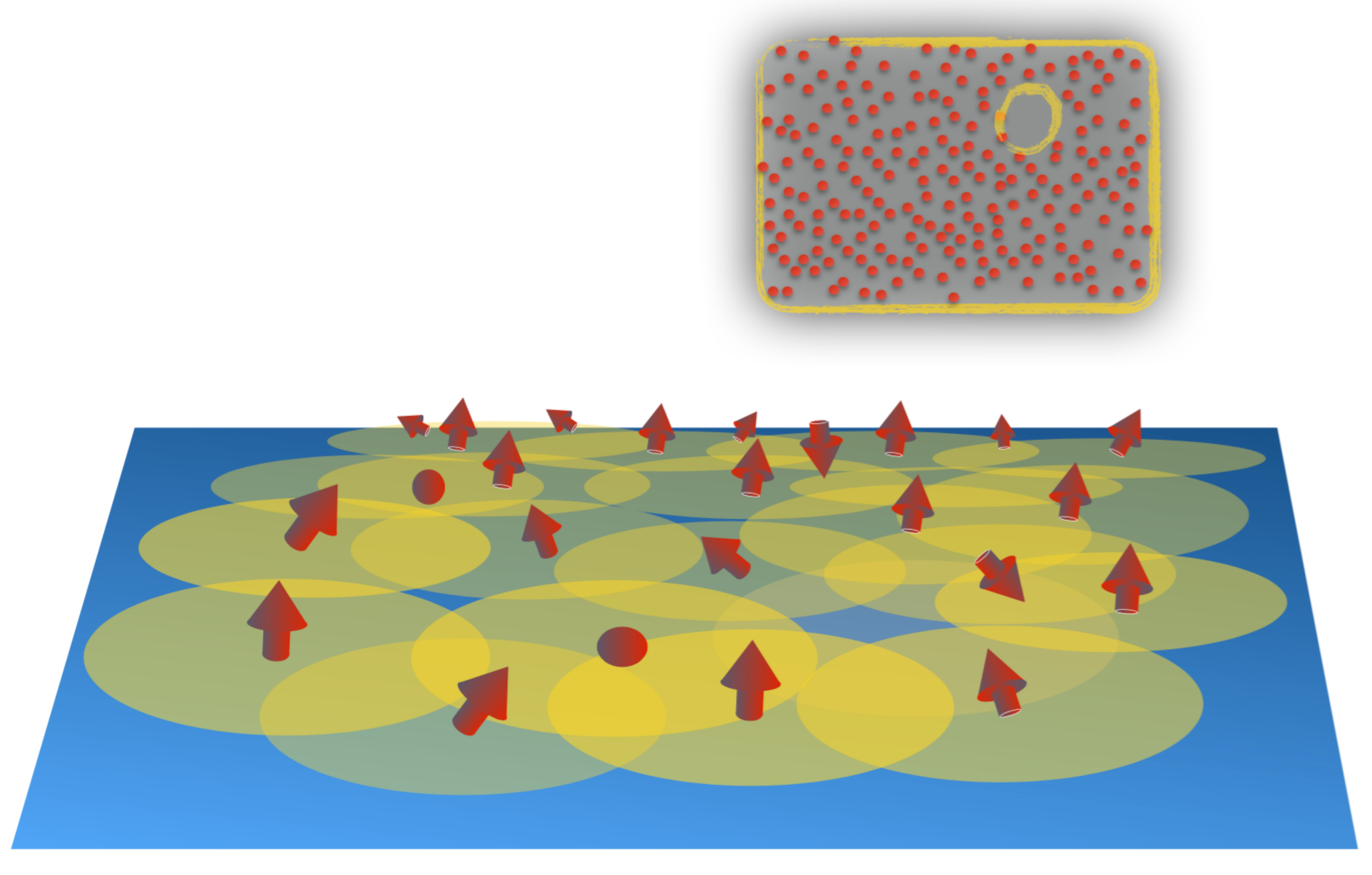}
\caption{ Structure of the Shiba glass. Magnetic moments (represented by a red arrow) on a superconductor  bind a subgap Yu-Shiba-Rusinov state (represented by a yellow disc) centred on the moments. A Shiba glass results from a hybridization of individual bound states in a random spatial distribution of moments. The collective amorphous state supports topological superconductivity above a critical moment density at finite out-of-plane polarization. Inset: a finite sample is enclosed by a topological edge mode of chiral Majorana states. Rare fluctuations give rise to antipuddles that exhibit localized low-energy excitations within a mobility gap protecting the topological phase.  }\label{x}
\end{figure}
The role of spatial symmetries in topological materials raises the question of how much spatial order is necessary for topological states to persist. In addition to the fundamental interest, possible realizations have far-reaching practical implications. The search for topological states has already moved beyond the elements found in nature to artificial man-made structures such as Majorana wires \cite{oreg,lutchyn, mourik}. The wires have the advantage of potentially allowing topological computation, but to carry out this function they must be almost defect free, which illustrates a generic complication in top-down fabrication strategies. On the other hand, fabrication of topological matter with randomly distributed constituents, if possible, would avoid that obstacle and offer new opportunities. A recent discovery of a mechanical gyroscopic metamaterial \cite{mitchell}, albeit purely classical system, suggests that also amorphous topological quantum matter could be achievable. By studying the properties of long-range hopping toy models, Agarwala and Shenoy pointed out the mathematical possibility of topological states with randomly localized states \cite{shenoy}. In this work, we propose the \emph{Shiba glass}, depicted in Fig.~1, as a concrete physical realization of amorphous topological quantum matter. Remarkably, we discover that (i) for a finite out-of-plane polarization the system supports topological superconductivity above a critical density despite complete absence of spatial order (ii) the topological phase is extremely robust and protected by a mobility gap (iii) the topological phase supports edge modes, whose signatures can be observed in standard Scanning Tunneling Microscopy (STM) experiments. The Shiba glass is fundamentally different from disordered topological materials which rely on band structures and thus on the spatial order of clean systems.

\textit{Theoretical description---} The studied amorphous topological superconductor is comprised of randomly distributed magnetic moments on a superconducting surface with a Rashba spin-orbit coupling.  The moments can arise from magnetic atoms, molecules or nanoparticles. Regular 1D structures of this type have been predicted to host Majorana states \cite{choy,np,brau1,klin,vazifeh,pientka2,bry,zhang,west} with supporting experimental evidence \cite{np2,ruby,pawlak}. More recently, ferromagnetic 2D lattices have emerged as a promising platform for chiral superconductivity \cite{menard2,rachel} with a rich topological phase diagram \cite{ront1}.  Classical magnetic moments embedded in a gapped $s$-wave superconductor give rise to Yu-Shiba-Rusinov (YSR) subgap states \cite{balatsky}, localized subgap states which decay algebraically for distances smaller than the superconducting coherence length. In 2D superconductors, such as layered systems, thin films and surfaces, the decay of the wavefunctions from the deep-lying impurity has a functional form $e^{-r/\xi}/\sqrt{k_Fr}$, where $\xi$ and $k_F$ are the superconducting coherence length and the Fermi wave vector of the underlying bulk. The Shiba glass results from a hybridization of randomly distributed YSR states. To model the system, we consider deep-lying YSR states with energies $\epsilon_0$ located in the vicinity of the gap center $\epsilon_0/\Delta\ll1$, where $\Delta$ is the pairing gap in the bulk.  The energy of a single YSR state is given by $\epsilon_0= \Delta\frac{1-\alpha^2}{1+\alpha^2}$, where $\alpha = \pi JS\mathcal{N}$ is a dimensionless impurity strength, $J$ is the magnetic coupling, $S$ is the magnitude of the magnetic moment and  $\mathcal{N}$ is the spin-averaged density of states at the Fermi level.  The deep-impurity assumption translates to $|1-\alpha|\ll 1$ and the energy of an impurity state is given by $\epsilon_0 \approx \Delta(1-\alpha)$. As outlined in the Supplementary Information (SI), the low-energy properties of the coupled impurity moments are modelled by a tight-binding Bogoliubov-de Gennes (BdG) Hamiltonian \cite{ront1}
\begin{equation}\label{h1}
\begin{gathered}
H_{mn} = \begin{pmatrix} h_{mn} & \Delta_{mn} \\ (\Delta_{nm})^* & -h_{mn}^* \end{pmatrix} ,
\end{gathered}
\end{equation}
which describes a long-range hopping between YSR states centred at random positions $\vec{r}_n$. The entries $h_{mn}$, $\Delta_{mn}$ for arbitrary configuration of magnetic moments is lengthy and given in the SI. Physical intuition can be obtained by considering the special case of fully out-of-plane ferromagnetic spins, where the model reduces to 
\begin{equation*}
\begin{split}
h_{mn} &= 
\Bigg\{\!\!\begin{array}{cl}
\epsilon_0 & m=n
\\
\displaystyle
\frac{\alpha\Delta}{2} \big[ I_1^-(r_{mn}) + I_1^+(r_{mn}) \big] & m\neq n
\end{array} 
\\
\Delta_{mn} &= 
\Bigg\{\!\!\begin{array}{cl}
0 & m=n
\\
\displaystyle
\frac{\alpha\Delta}{2}  \big[ I_4^+(r_{mn}) - I_4^-(r_{mn}) \big] \frac{x_{mn}\! - iy_{mn}}{r_{mn}} & m\neq n.
\end{array}
\end{split}
\end{equation*}
In the above expression $r_{mn}=|\vec{r}_m-\vec{r}_n|$, and $x_{mn}$ and $y_{mn}$ are components of $\vec{r}_m-\vec{r}_n\equiv(x_{mn},y_{mn})$.  The hopping elements are expressed in terms of the functions
\begin{equation*}
\begin{split}
&I_4^\pm(r) = \frac{\mathcal{N}_\pm}{\mathcal{N}}\, \Re \Big[ iJ_{1}\big(k_F^\pm r +ir/\xi\big) + H_{-1}\big(k_F^\pm r +ir/\xi\big) \Big] \\
&I_1^\pm(r) = \frac{\mathcal{N}_\pm}{\mathcal{N}} 
\Re \Big[ J_0\big( k_F^\pm r +ir/\xi \big) +iH_0\big( k_F^\pm r +ir/\xi \big) \Big],
\end{split}
\end{equation*}
where $J_n$ and $H_n$ are Bessel and Struve functions of order $n$. The Rashba spin-orbit coupling induces two helical Fermi surfaces with density of states $\mathcal{N}_\pm=\mathcal{N}(1\mp\lambda/\sqrt{1+\lambda^2})$ and Fermi wavenumber $k_F^\pm = k_F(\sqrt{1 + \lambda^2} \mp \lambda)$, where $\lambda = \alpha_R/(\hbar v_F)$ is the dimensionless Rashba coupling and $k_F, v_F$ the Fermi wavenumber and velocity in the absence of spin-orbit coupling. The Rashba coupling also slightly modifies the superconducting coherence length $\xi = (\hbar v_F/\Delta) \sqrt{1 + \lambda^2}$. For ferromagnetic textures, the pairing term $\Delta_{ij}$ vanishes  with vanishing Rashba coupling $\alpha_R=0$. The low-energy Hamiltonian (\ref{h1}) describes an odd-parity pairing $\Delta_{mn}=-\Delta_{nm}$ which is a long-range hopping variant of a $p_x+ip_y$ superconductivity.  In  Eq.~(\ref{h1}) the hopping and pairing functions decay as $f(r)\propto\frac{e^{-r/\xi}}{r^{1/2}}$ and display oscillations at wave vectors $k_F^\pm$.

\begin{figure*}
\includegraphics[width=0.85\linewidth]{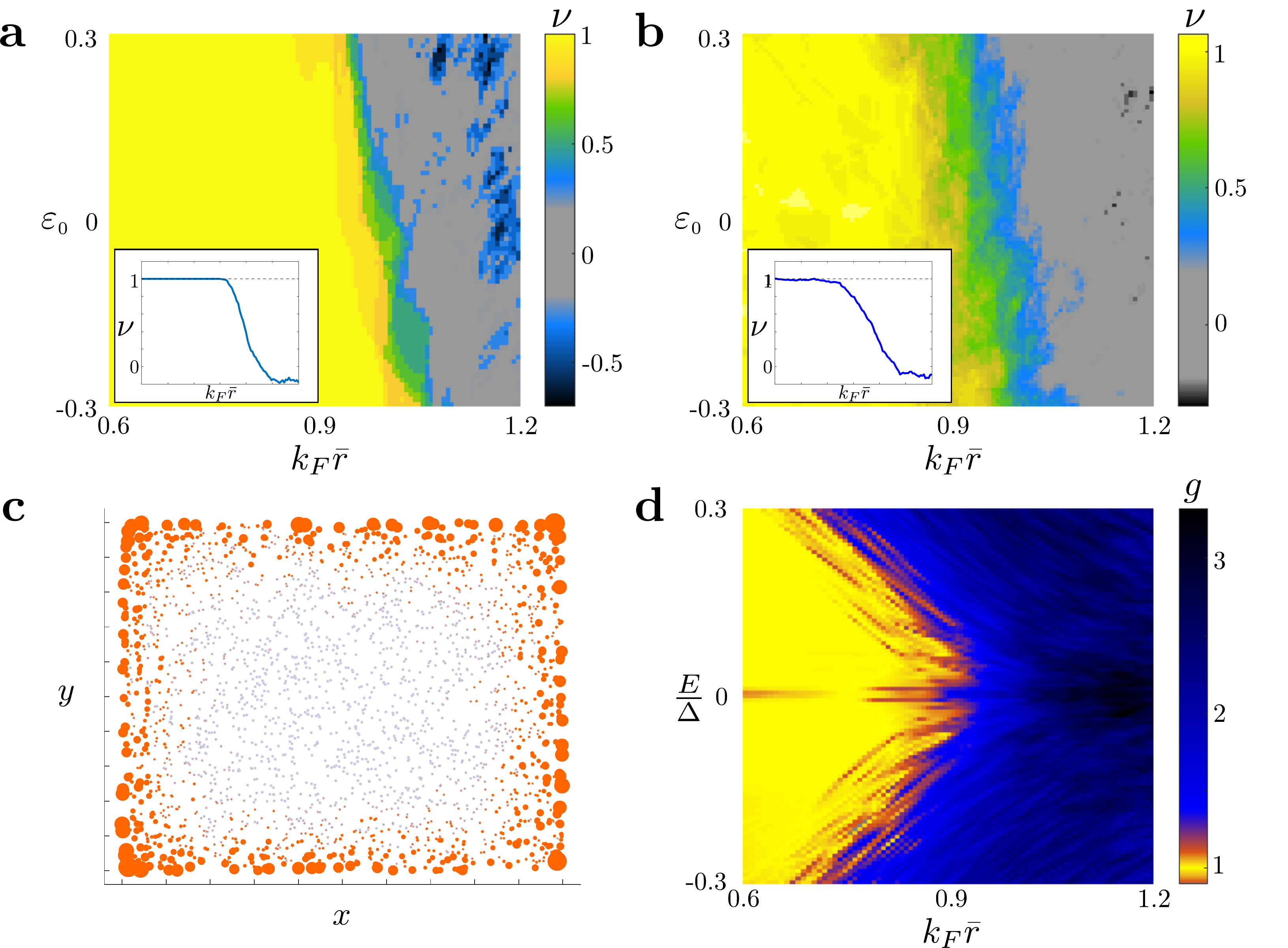}
\caption{ Topological superconductivity in the Shiba glass.
\textbf{a}, Topological phase diagram for a ferromagnetic Shiba glass as a function of the single-moment bound-state energy $\epsilon_0$ and the characteristic length between the moments $\bar r = \rho^{-\frac{1}{2}}$, where $\rho$ is the moment density per unit area. The colour bar indicates the value of the Chern number. The adatom number is held fixed at 600, with $k_F \xi = \frac{4\pi}{5}$ and $\lambda = 0.2$. The displayed diagram is an average over 10 configurations. The high Rashba splitting $\lambda$ and low value of $k_F\xi$ are appropriate for a proximity superconducting 2D semiconductor; phase diagrams for parameters more appropriate for metals are presented in the SI.
Inset: Line along $\epsilon_0 = 0.1$, averaged over 500 configurations.  \textbf{b,} Same as in \textbf{a}, but for magnetic moment directions drawn from the Boltzmann distribution with $\beta E_z = 10$ and averaged over 30 configurations, and with the number of moments fixed at 900. The deviation from the quantized values and the width of the transition region diminish as the system size is increased.  \textbf{c,} Local density of states for a $12.5\xi \times 12.5\xi$ square Shiba glass system comprising $2500$ randomly distributed  sites, integrated over subgap energies $|E| < 0.1\Delta$. Parameters used same as in \textbf{a}, with onsite energy $\varepsilon_0 = 0$. The areas of the orange discs correspond to the magnitude of the LDOS; each site is additionally represented by a gray point which is visible when the LDOS is negligible.
\textbf{d,} The thermal conductance (in units of $\frac{\pi k_B^2T}{3\hbar}$)  along the line $\epsilon_0 = 0$ for the same system parameters as in the previous figures, but with 2500 adatoms. The vertical width of the conduction plateau (yellow) corresponds to the mobility gap of the system, and can be seen to close as the system approaches the transition to the trivial gapless phase.}\label{Fig:diagrams}
\end{figure*}

\textit{Physical properties of the Shiba glass ---}  The spectrum and the topological phase diagram of a finite system can be calculated by diagonalizing the effective Hamiltonian \eqref{h1} for spatially uncorrelated random positions of magnetic moments. After deriving the finite-size properties, we discuss the extrapolation to the thermodynamic limit. For 2D time-reversal breaking topological superconductors the relevant topological index classifying the state is the Chern number. We will evaluate Chern numbers by employing a real-space approach \cite{zhang2} as explained in the SI.

By evaluating the Chern number, we uncover the topological phase diagram of finite Shiba glass systems which can be seen in Fig.~2a. For sufficiently high densities, a ferromagnetically ordered system is generally in a topological phase with Chern number $|\mathcal C| = 1$. For the employed parameters, the critical density $\rho_c$ corresponds to the characteristic length scale $\bar{r}_c=\rho_c^{-1/2}\approx k_F^{-1}$. For lower densities $(\bar{r} \gg k_F^{-1})$, the system is in general topologically trivial and gapless; rare configurations can manage to enter a topological phase but do not survive disorder averaging. The pattern persists even when the directions of the local spins deviate from the perfect ferromagnetic configuration; in Fig.~\ref{Fig:diagrams}b we plot the phase diagram for spin configurations drawn from a thermal distribution where the angles $\theta_j$ between the moments and the surface normal  are determined by the Boltzmann weights  $e^{-\beta E_Z \cos\theta_j}$. This situation corresponds to an ensemble of decoupled spins at Zeeman field $E_Z$ polarizing the moments perpendicular to the plane and disordered by thermal fluctuations at inverse temperature $\beta$. Alternatively, the situation can be regarded as a magnetic disorder where the disorder is parametrized by the thermal distribution and $\beta E_Z$ instead of some other random distribution. For $\beta E_Z= 10$, as indicated by Fig.~\ref{Fig:diagrams}b, the phase diagram remains qualitatively unchanged when compared to that for the completely polarized case. The robustness to moment disorder is not an artefact of the thermal distribution, and we discover qualitatively similar results for other disorder averages exhibiting comparable polarization.   

The physical consequences of the topological nature of the Shiba glass are illustrated in Figs.~2c and 2d. The first one shows that the local density of states (LDOS) is concentrated on the sample edges. This is a consequence of a topological edge mode enclosing a finite system and is directly observable as discussed below. In Fig.~\ref{Fig:diagrams}d we have plotted the thermal conductance of finite systems coupled to external leads (see the SI for details). In the topological phase, the system exhibits a quantized thermal conductance which is a direct consequence of the nontrivial topology. The quantized conductance is effected by the edge modes despite the system being highly irregular in real space. In finite-size systems, for parameters close to the phase boundary, the quantized conductance plateau is destroyed and the conductance assumes continuous values. The non-quantized conductance in the trivial phase indicates the low-energy states there extend over the sample.   

Now we turn to discuss the features seen when increasing the system size. First of all, in the thermodynamic limit the Shiba glass phase is gapless. While this is a generic feature of a superconductor with magnetic impurities \cite{balatsky}, a qualitatively new mechanism for low-energy excitations arises in the topological phase. These emerge from rare fluctuations that leave a substantial area where magnetic moments are sparse.  As depicted in Fig.~1, these empty \emph{antipuddles} give rise to low-energy modes which are reminiscent of the gapless edge states circulating around a hole punched in a gapped topological phase. While the probability of formation of antipuddles is exponentially suppressed as a function of their size and their effect is relatively unimportant in finite systems with high density, in infinite systems antipuddles give rise to a tail down to zero energy in the DOS.  The antipuddle mechanism provides a simple physical argument why the energy gap must scale to zero in the thermodynamic limit. The second important notion is that, in the thermodynamic limit, the system has well-defined topological nature despite being gapless. The low-energy modes, as we have argued above, are localized perturbations and the states  with non-localized wavefunctions have a finite energy threshold. Thus, instead of an energy gap, the system exhibits a  mobility gap protecting the topological state. This behaviour is analogous to the integer quantum Hall effect where the extended states carrying Chern numbers are separated by localized states in the Landau level gap \cite{prange}. In the SI we have calculated the thermal conductance for an antipuddle configuration, which shows that for isolated antipuddles, the system has a vanishing energy gap but a finite well-defined mobility gap within which the heat conductance is quantized. In the topological phase the antipuddles are rare and effectively decoupled, thus they cannot destroy the conductance quantization.  

\textit{Discussion ---} In our work we have not addressed the question of the magnetic ordering but rather shown that a finite polarization perpendicular to the plane gives rise to a topological phase. The nature of the ordering would likely depend sensitively on the specific physical realization; however, there exist a number of mechanisms driving the system to a polarized state. For example, ignoring the modifications arising from superconductivity at large distances \cite{yao}, the RKKY coupling leads to an effective interaction $H=\sum_{i\neq j} J(|\mathbf{r}_i-\mathbf{r}_j|)\mathbf{S}_i\cdot \mathbf{S}_j$ between the moments, where the sign of  $J(r)$ oscillates as a function of the position as $\cos(2k_Fr)$. While for distances $2k_Fr\gg1$ this leads to complicated frustrated behaviour, in sufficiently dense systems  $\pi/4<k_Fr<3\pi/4$ the interaction is effectively ferromagnetic. Therefore, in the large part of the topological region $k_F\bar{r}\lesssim 1$, this mechanism favours a ferromagnetic ordering polarizing the system. In addition, an anisotropic crystal field splitting $DS_z^2$ and an external Zeeman field $B S_z$ would drive the system towards an out-of-plane polarization. 

The studied  Shiba glass  system could be realized by decorating an effective 2D or a layered 3D superconductor with magnetic atoms or molecules. Considering the requirement $k_F\bar{r}\lesssim 1$, dilute electron systems such as proximity-superconducting 2D semiconductors with Rashba spin-orbit coupling are promising candidate systems. Another candidate system is the layered superconductor NbSe$_2$ where 2D YSR states \cite{menard1} and their coupling have been observed \cite{kezilebieke} recently. The most direct experimental probe is provided by STM measurement of the LDOS. As shown above, in the topological phase the Shiba glass system exhibits a significant concentration of the subgap LDOS at the sample boundaries which can be directly observed by STM. This signal is clearly detectable at temperatures below the mobility gap scale which can be of the order of $k_BT=0.1\Delta-0.3\Delta$ as shown in Fig.~2 d. 

In summary, we introduced the Shiba glass as a platform for amorphous topological superconductivity and elucidated the general properties of such systems. Our results illustrate the physical feasibility of amorphous topological quantum matterials and provide a concrete prescription to experimentally realize and observe them. Our discovery motivates expanding the search for topological materials beyond crystalline systems and paves the way for fabricating topological matter from nontopological materials with random dopants.

\textbf{Acknowledgements ---\hspace{3mm}} This work is supported by The Academy of Finland (T.O.), the Aalto Centre for Quantum
Engineering (T.O.), the Swedish Cultural Foundation in Finland (K.P.) and the Vilho, Yrjö and Kalle Väisälä Foundation of the Finnish Academy of Science and Letters (A.W.).

%
%
%
%
%
%

\begin{widetext}

\begin{center}
\large Supplementary Information
\end{center}

\section{Microscopic derivation of the model}
In this section, we present a microscopic derivation of the effective Shiba glass Hamiltonian studied in the main text. The derivation follows the spirit of Ref.~\onlinecite{pientka2}, but proceeds differently since we are considering 2D systems with a spin-orbit coupling and arbitrary spin textures rather than 1D helical magnetic chains. Similar 1D\cite{bry} and 2D\cite{ront1,ront2} systems have been previously studied on regular lattices with ferromagnetic textures where significant simplification occurs. Our starting point is the mean-field Bogoliubov-de Gennes  equation for magnetic impurities on a 2D $s$-wave superconductor with Rashba spin-orbit coupling,
\begin{equation}
\left[\xi_k \tau_z + \alpha_R(\bs\sigma\times\vec k)_z\tau_z + \Delta\tau_x + JS\sum_j (\vec{\hat S}_j\cdot\bs\sigma) \delta(\vec r - \vec r_j)\right]\Psi = E\Psi
\end{equation}
where $\xi_k = \frac{k^2}{2m} - \mu$ represents the kinetic energy and chemical potential, $\alpha_R$ the spin-orbit coupling, $\Delta$ the superconducting pairing amplitude, and $J$ and $S$ are the exchange coupling and the magnitude of the classical impurity spin, respectively. The arbitrary positions of $N$  magnetic moments are labelled by $\vec r_j$ and they point in the direction defined by the unit vectors $\hat S_j$.  The $\tau$ and $\sigma$ matrices correspond to Pauli matrices in particle-hole and spin subspaces; the vector basis is
$\Psi = (\psi_\uparrow,\psi_\downarrow,\psi^\dagger_\downarrow,-\psi^\dagger_\uparrow)^T$. By Fourier transforming, the equation can be re-expressed as 
\begin{equation}
\Psi(\vec r_i) = JS \sum_j\int \frac{d\vec k}{(2\pi)^2} e^{i\vec k \cdot \vec r_{ij}}G_0(\vec k) (\vec{\hat S_j} \cdot \sigma) \Psi(\vec r_j),
\end{equation}
with $\vec r_{ij} = \vec r_i - \vec r_j$, and where $G_0(\vec k)$ is the Green's function of the superconductor without impurities. Separating the terms dependent on $\vec r_i$, the above can be written as
\begin{equation}
\left[\vec{\hat S_i} \cdot \bs\sigma - \alpha\frac{E + \Delta \tau_x}{\sqrt{\Delta^2 - E^2}}\right]\Psi(\vec r_i) = -\sum_{j\neq i} (\vec{\hat S}_i \cdot \bs \sigma) J_E(\vec r_{ij}) (\vec{\hat S}_j \cdot \bs \sigma) \Psi(\vec r_j),\label{app:eq:JEintro}
\end{equation}
where $\alpha = \frac{1}{2}JSm$ and
\begin{equation}
J_E(\vec r) = \frac{JS}{2}\int \frac{d\vec k}{(2\pi)^2} e^{i\vec k \cdot \vec r}\left[G_0^+(\vec k) + G_0^-(\vec k)\right].
\end{equation}
Due to the Rashba coupling, the substrate Green's function splits into two helical components  
\begin{equation}
G_0^\nu = \frac{E + \xi_\nu \tau_z + \Delta \tau_x}{E^2 - \xi_\nu^2 - \Delta^2}\left[1 + \nu(\frac{k_y}{k}\sigma_x - \frac{k_x}{k}\sigma_y)\right],
\end{equation}
where the helical branches $\nu=\pm 1$ have dispersions $\xi_\nu = \xi_k + \nu \alpha_R k$. Evaluation of the Green's function can be carried out by linearizing at the Fermi surface $\xi_\nu = 0$ which yields
\begin{equation}
J_E(\vec r) \approx \frac{1}{2}JSm\sum_{\nu = \pm 1} (1 - \nu \frac{\lambda}{\sqrt{1 + \lambda^2}})\int \frac{d\xi}{2\pi}\int \frac{d\theta}{2\pi}e^{ik_\nu r \cos(\phi - \theta)}\frac{E + \xi_\nu \tau_z + \Delta \tau_x}{E^2 - \xi_\nu^2 - \Delta^2}\left[1 + \nu(\sin\theta\sigma_x - \cos\theta\sigma_y)\right],
\end{equation}
where $\phi$ is the angle formed with $\vec r$ and the $x$ axis. Hence, we can write
\begin{equation}
\begin{split}
J_E(\vec r_{ij}) = \frac{\alpha}{2}\bigg[&(E + \Delta\tau_x)(I_1^+(r_{ij}) + I_1^-(r_{ij})) - i\tau_z(\frac{y_i-y_j}{r_{ij}}\sigma_x - \frac{x_i-x_j}{r_{ij}}\sigma_y)(I_0^+(r_{ij}) - I_0^-(r_{ij}))\\
& + \tau_z(I_2^+(r_{ij}) + I_2^-(r_{ij})) + (E + \Delta\tau_x)(\frac{y_i-y_j}{r_{ij}}\sigma_x - \frac{x_i-x_j}{r_{ij}}\sigma_y)(I_3^+(r_{ij}) - I_3^-(r_{ij}))\bigg],
\end{split}
\end{equation}
where we have defined the integrals
\begin{align*}
I_0^\pm(r) &= i\frac{N_\pm}{2\pi^2}\int_{-\infty}^\infty d\xi \int_0^{2\pi}d\phi \frac{\xi e^{ik^\pm r\cos\phi + i\phi}}{E^2 - \xi^2 - \Delta^2}  = N_\pm\RE\left[iJ_1(k_F^\pm r + i\frac{r}{\xi_E}) + \frac{2}{\pi} - H_1(k_F^\pm r + i\frac{r}{\xi_E})\right]\\
I_1^\pm(r) &= \frac{N_\pm}{2\pi^2}\int_{-\infty}^\infty d\xi \int_0^{2\pi}d\phi \frac{e^{ik^\pm(\xi) r\cos\phi}}{E^2 - \xi^2 - \Delta^2}  = \frac{-N_\pm}{\sqrt{\Delta^2 - E^2}} \RE\left[J_0(k_F^\pm r + i\frac{r}{\xi_E}) + iH_0(k_F^\pm r + i\frac{r}{\xi_E})\right]\\
I_2^\pm(r) &= \frac{N_\pm}{2\pi^2}\int_{-\infty}^\infty d\xi \int_0^{2\pi}d\phi \frac{\xi e^{ik^\pm(\xi) r\cos\phi}}{E^2 - \xi^2 - \Delta^2}  = N_\pm\IM\left[J_0(k_F^\pm r + i\frac{r}{\xi_E}) + iH_0(k_F^\pm r + i\frac{r}{\xi_E})\right]\\
I_3^\pm(r) &= \frac{N_\pm}{2\pi^2}\int_{-\infty}^\infty d\xi \int_0^{2\pi}d\phi \frac{e^{ik^\pm(\xi) r\cos\phi + i\phi}}{E^2 - \xi^2 - \Delta^2}  = -i\frac{N_\pm}{\sqrt{\Delta^2 - E^2}} \IM\left[iJ_1(k_F^\pm r + i\frac{r}{\xi_E}) + \frac{2}{\pi} - H_1(k_F^\pm r + i\frac{r}{\xi_E})\right].
\end{align*}
In the above, we have used the dimensionless spin-orbit coupling $\lambda = \frac{\alpha_R}{v_F}$ and the superconducting coherence length $\xi_E = \frac{v_F}{\sqrt{\Delta^2-E^2}}$  as well as the densities of states of the two helical branches  $N_\pm = 1 \mp \frac{\lambda}{\sqrt{1 + \lambda^2}}$, their linearized dispersions  $k^\pm(\xi) = \frac{\xi}{v_F} + k_F^\pm $, as well as their Fermi wave vectors $k_F^\pm = k_F(\sqrt{1 + \lambda^2} \mp \lambda)$. In general, the integrals above decay with site distance as $f(r) \propto e^{-r/\xi_E}/\sqrt{r}$; however, $I_0$ and $I_2$ contain terms decaying asymptotically as $r^{-2}$ and $ r^{-1}$, respectively.

We now introduce the quantities
\begin{align}
S_{ij} &= \sum_{\nu}  N_\nu \left[J_0(k_F^\nu r_{ij} + i\frac{r_{ij}}{\xi_E}) + iH_0(k_F^\nu r_{ij} + i\frac{r_{ij}}{\xi_E})\right]\\
A_{ij} &= \sum_{\nu}  \nu N_\nu \left[iJ_1(k_F^\pm r_{ij} + i\frac{r_{ij}}{\xi_E}) + \frac{2}{\pi} - H_1(k_F^\pm r_{ij} + i\frac{r_{ij}}{\xi_E})\right]
\end{align}
in terms of which the integrals can be written
\begin{align}
\RE(S_{ij}) &= -\sqrt{\Delta^2 - E^2}(I_1^+(\vec r_{ij}) + I_1^-(\vec r_{ij}))\\
\IM(S_{ij}) &= I_2^+(\vec r_{ij}) + I_2^-(\vec r_{ij})\\
\RE(A_{ij}) &= I_0^+(\vec r_{ij}) - I_0^-(\vec r_{ij})\\
i\IM(A_{ij}) &= -\sqrt{\Delta^2 - E^2}(I_3^+(\vec r_{ij}) - I_3^-(\vec r_{ij})).
\end{align}
To simplify notation, we define the operator $\kappa_{ij} = \frac{y_i-y_j}{r_{ij}}\sigma_x - \frac{x_i-x_j}{r_{ij}}\sigma_y$. Eq.~\eqref{app:eq:JEintro} now takes the form
\begin{equation}\label{intermediate}
\begin{split}
\left[\vec{\hat S}_i \cdot \bs\sigma - \alpha\frac{E + \Delta \tau_x}{\sqrt{\Delta^2 - E^2}}\right]\Psi(\vec r_i) = -\frac{\alpha}{2}\sum_{j\neq i} (\vec{\hat S}_i \cdot \bs \sigma)\bigg[&-\frac{E + \Delta\tau_x}{\sqrt{\Delta^2-E^2}}\RE S_{ij} - i\tau_z\kappa_{ij}\RE A_{ij}\\
& + \tau_z \IM S_{ij} -i \frac{E + \Delta\tau_x}{\Delta^2 - E^2}\kappa_{ij}\IM A_{ij}  \bigg](\vec{\hat S}_j \cdot \bs \sigma)\Psi(\vec r_j).
\end{split}
\end{equation}
The next step is transforming Eq.~\eqref{app:eq:JEintro} to the local BdG basis $\lbrace \ket{+}_{\tau_x}\otimes\ket{\uparrow_j}, \ket{-}_{\tau_x}\otimes\ket{\downarrow_j}, \ket{+}_{\tau_x}\otimes\ket{\downarrow_j}, \ket{-}_{\tau_x}\otimes\ket{\uparrow_j}\rbrace$. The spin states are defined in terms of the spin directions $\vec{\hat S}_j = (\sin\theta_j\cos\phi_j,\sin\theta_j\sin\phi_j,\cos\theta_j)$ as 
\begin{equation}
\ket{\uparrow_j} = \frac{1}{\sqrt{N}} \begin{pmatrix}
e^{-i\frac{\phi_j}{2}}\cos\frac{\theta_j}{2}\\
e^{i\frac{\phi_j}{2}}\sin\frac{\theta_j}{2}
\end{pmatrix}
\qquad
\ket{\downarrow_j} = \frac{1}{\sqrt{N}} \begin{pmatrix}
-e^{-i\frac{\phi_j}{2}}\sin\frac{\theta_j}{2}\\
e^{i\frac{\phi_j}{2}}\cos\frac{\theta_j}{2}.
\end{pmatrix},
\end{equation}
and their matrix elements are
\begin{align}
\brakets{\uparrow_i}{\uparrow_j} &= \cos\frac{\theta_i}{2}\cos\frac{\theta_j}{2}e^{i\frac{\phi_i-\phi_j}{2}} + \sin\frac{\theta_i}{2}\sin\frac{\theta_j}{2}e^{-i\frac{\phi_i-\phi_j}{2}}\nonumber\\
\brakets{\uparrow_i}{\downarrow_j} &= -\cos\frac{\theta_i}{2}\sin\frac{\theta_j}{2}e^{i\frac{\phi_i-\phi_j}{2}} + \sin\frac{\theta_i}{2}\cos\frac{\theta_j}{2}e^{-i\frac{\phi_i-\phi_j}{2}}\nonumber\\
\bra{\uparrow_i}\sigma_x\ket{\uparrow_j} &=
\cos\frac{\theta_i}{2}\sin\frac{\theta_j}{2}e^{i\frac{\phi_i+\phi_j}{2}} + \sin\frac{\theta_i}{2}\cos\frac{\theta_j}{2}e^{-i\frac{\phi_i+\phi_j}{2}}\nonumber\\
\bra{\uparrow_i}\sigma_y\ket{\uparrow_j} &=
-i\cos\frac{\theta_i}{2}\sin\frac{\theta_j}{2}e^{i\frac{\phi_i+\phi_j}{2}} + i\sin\frac{\theta_i}{2}\cos\frac{\theta_j}{2}e^{-i\frac{\phi_i+\phi_j}{2}}\nonumber\\
\bra{\uparrow_i}\sigma_x\ket{\downarrow_j} &=
\cos\frac{\theta_i}{2}\cos\frac{\theta_j}{2}e^{i\frac{\phi_i+\phi_j}{2}} - \sin\frac{\theta_i}{2}\sin\frac{\theta_j}{2}e^{-i\frac{\phi_i+\phi_j}{2}}\nonumber\\
\bra{\uparrow_i}\sigma_y\ket{\downarrow_j} &=
-i\cos\frac{\theta_i}{2}\cos\frac{\theta_j}{2}e^{i\frac{\phi_i+\phi_j}{2}} - i\sin\frac{\theta_i}{2}\sin\frac{\theta_j}{2}e^{-i\frac{\phi_i+\phi_j}{2}}.\nonumber
\end{align}

For notational convenience, we introduce the matrices
\begin{align}
a^{\sigma \sigma'}_{ij} &= \frac{1}{2}\brakets{\sigma_i}{\sigma_j'} \RE S_{ij}\\
b^{\sigma \sigma'}_{ij} &= \frac{1}{2}\bra{\sigma_i}\hat\kappa_{ij}\ket{\sigma_j'} \RE A_{ij}\\
c^{\sigma \sigma'}_{ij} &= \frac{1}{2}\bra{\sigma_i}\hat\kappa_{ij}\ket{\sigma_j'} \IM A_{ij}\\
d^{\sigma \sigma'}_{ij} &= \frac{1}{2}\brakets{\sigma_i}{\sigma_j'} \IM S_{ij},
\end{align} 
where $|\sigma_j\rangle$ can be either $|\uparrow_j\rangle$ or$|\downarrow_j\rangle$. Now we can write Eq.~\eqref{intermediate} in the $4N\times 4N$ form, defining $u \equiv (\Delta + E)/\sqrt{\Delta^2-E^2}$:
\begin{equation}
\begin{split}
\begin{pmatrix}
1 - \alpha u& 0 & 0 & 0\\
0 & -1 + \alpha u^{-1}& 0 & 0\\
0 & 0 & -1 - \alpha u & 0\\
0 & 0 & 0 & 1 + \alpha u^{-1}
\end{pmatrix}
\Psi =
-\alpha&\left[
\begin{pmatrix}
- u a^{\uparrow \uparrow} & 0 & - u a^{\uparrow \downarrow} & 0\\
0 &  u^{-1}a^{\downarrow \downarrow} & 0 &  u^{-1}a^{\downarrow \uparrow}\\
- u a^{\downarrow \uparrow} & 0 & - u a^{\downarrow \downarrow} & 0\\
0 &  u^{-1}a^{\uparrow \downarrow} & 0 &  u^{-1}a^{\uparrow \uparrow} 
\end{pmatrix}\right.
\\
&+ \begin{pmatrix}
0 & -ib^{\uparrow \downarrow} & 0 & -ib^{\uparrow \uparrow}\\
-ib^{\downarrow \uparrow} & 0 & -ib^{\downarrow \downarrow} & 0\\
0 & -ib^{\downarrow \downarrow}& 0 & -ib^{\downarrow \uparrow}\\
-ib^{\uparrow \uparrow} & 0 & -ib^{\uparrow \downarrow} & 0
\end{pmatrix}
+
\begin{pmatrix}
0 & d^{\uparrow \downarrow} & 0 & d^{\uparrow \uparrow}\\
d^{\downarrow \uparrow} & 0 & d^{\downarrow \downarrow} & 0\\
0 & d^{\downarrow \downarrow} & 0 & d^{\downarrow \uparrow}\\
d^{\uparrow \uparrow} & 0 & d^{\uparrow \downarrow} & 0
\end{pmatrix}
\\
&+
\left.\begin{pmatrix}
 -i u c^{\uparrow \uparrow} & 0 & -i u c^{\uparrow \downarrow} & 0\\
 0 &  i u^{-1} c^{\downarrow \downarrow} & 0 & i u^{-1} c^{\downarrow \uparrow}\\
-i u c^{\downarrow \uparrow} & 0 & -i u c^{\downarrow \downarrow} & 0\\
0 & i u^{-1} c^{\uparrow \downarrow} & 0 & i u^{-1} c^{\uparrow \uparrow}
\end{pmatrix}
\right]\Psi.
\end{split}
\end{equation}
Up to this point, the only approximation has been the standard evaluation of Green's functions by linearizing the dispersion at the Fermi level. Now we concentrate on the low energy behaviour in the vicinity of the gap center $|E| \ll \Delta$ and work in the linear order in $E/\Delta$. We now project the problem to the basis of subgap Shiba states,\cite{pientka2,bry,ront1,ront2} which corresponds to the upper-left $2N\times 2N$ block:
\begin{equation}
\begin{split}
\begin{pmatrix}
1 - \alpha u& 0\\
0 & -1 + \alpha u^{-1}
\end{pmatrix}
\Psi =
\alpha&\Bigg[
\begin{pmatrix}
 u a^{\uparrow \uparrow} & 0\\
0 & - u^{-1}a^{\downarrow \downarrow}
\end{pmatrix}
+
\begin{pmatrix}
 i u c^{\uparrow \uparrow} & 0 \\
 0 &  -i u^{-1} c^{\downarrow \downarrow}
\end{pmatrix}
\\
&+ \begin{pmatrix}
0 & ib^{\uparrow \downarrow}\\
ib^{\downarrow \uparrow} & 0
\end{pmatrix}
+
\begin{pmatrix}
0 & -d^{\uparrow \downarrow}\\
-d^{\downarrow \uparrow} & 0
\end{pmatrix}
\Bigg]\Psi.
\end{split}
\end{equation}
In addition, we assume that the YSR states with energies $\epsilon_0=\frac{1-\alpha^2}{1+\alpha^2}\Delta$ determined by the dimensionless Shiba coupling $\alpha$ lie close to the gap center, so that $|1 - \alpha| \ll 1$. 
Neglecting terms quadratic in these, we get 
\begin{equation}
\begin{split}
\begin{pmatrix}
1 - \alpha - \frac{E}{\Delta}& 0\\
0 & -1 + \alpha - \frac{E}{\Delta}
\end{pmatrix}
\Psi =
&\Bigg[
\frac{E}{\Delta}
\begin{pmatrix}
a^{\uparrow \uparrow} + i c^{\uparrow \uparrow}& 0\\
0 & a^{\downarrow \downarrow} +  i c^{\downarrow \downarrow}
\end{pmatrix}\\
&+
\alpha
\begin{pmatrix}
a^{\uparrow \uparrow} +i c^{\uparrow \uparrow}& 0\\
0 & -a^{\downarrow \downarrow} +  -i c^{\downarrow \downarrow}
\end{pmatrix}
+ \alpha\begin{pmatrix}
0 & -d^{\uparrow \downarrow}+ib^{\uparrow \downarrow}\\
-d^{\downarrow \uparrow}+ib^{\downarrow \uparrow} & 0
\end{pmatrix}
\Bigg]\Psi,
\end{split}
\end{equation}
where the submatrices $a,b,c,d$ are to be evaluated at $E = 0$. We isolate the energy:
\begin{equation}\label{penultimate}
\begin{split}
\frac{E}{\Delta}
\begin{pmatrix}
1 + a^{\uparrow \uparrow} + i c^{\uparrow \uparrow} & 0 \\
0 & 1 + a^{\downarrow \downarrow} + i c^{\downarrow \downarrow}
\end{pmatrix}\Psi
&= 
\alpha
\begin{pmatrix}
(1 - \alpha)  - a^{\uparrow \uparrow}  - i c^{\uparrow \uparrow}& d^{\uparrow \downarrow}-ib^{\uparrow \downarrow}\\
d^{\downarrow \uparrow}-ib^{\downarrow \uparrow} & -(1 - \alpha) +a^{\downarrow \downarrow} + i c^{\downarrow \downarrow}
\end{pmatrix}\Psi
\end{split}.
\end{equation}
This equation can be written in the form $H\Psi=E\Psi$ in terms of an effective Hamiltonian
\begin{equation}\label{app:eq:herm}
	H_{ij} =
	\begin{pmatrix}
		h_{ij} & \Delta_{ij} \\
		\Delta^*_{ji} & -h^*_{ij}
	\end{pmatrix},
\end{equation}
where
\begin{align}
h_{ij} &= 
\epsilon_0\delta_{ij} + \alpha\Delta(a^{\uparrow \uparrow}_{ij} + i c^{\uparrow \uparrow}_{ij})\\
\Delta_{ij} &= \alpha(d^{\uparrow \downarrow}_{ij}-ib^{\uparrow \downarrow}_{ij}).
\end{align}
In writing Eq.~\eqref{penultimate} to the form \eqref{app:eq:herm}, we have omitted elements $\propto E(a^{\uparrow \uparrow} + i c^{\uparrow \uparrow})$ since these give insignificant contribution to energies when $|E|/\Delta\ll 1$. The reason is that in the low-energy regime, the expectation value of $(a^{\uparrow \uparrow} + i c^{\uparrow \uparrow})$ turn out to be of the order of $E$, and the overall contribution is of the order $\mathcal{O}((E/\Delta)^2)$ as can be verified by perturbation theory. Thus the effective Hamiltonian \eqref{app:eq:herm}  reduces to the one quoted in the main text and captures the low-energy dynamics accurately to linear order in $E/\Delta$.   The accuracy of the model can also be checked by numerically comparing the low-energy spectra of Eq.~\eqref{app:eq:herm} to that obtained without the approximation and we find excellent agreement in the studied regime.

\section{Real-space evaluation of the topological invariant}
To find the topological phase diagram, we need to evaluate topological invariants in real space. The relevant topological index for 2D systems with broken time-reversal symmetry is the Chern number. This is generally obtained in $\vec k$-space, but there are various methods of computing it in real space as well.\cite{loring,zhang2} A comparison shows that these methods are generally of similar computational efficiency and yield the same values for the topological invariant.

The real-space Chern number method of Ref. \onlinecite{zhang2} proceeds by defining the coupling matrices $C_{\alpha,\alpha+1}$, with elements
\begin{equation}
C^{mn}_{\alpha,\alpha+1} = \bra{\psi^m}e^{i(\vec q_\alpha - \vec q_{\alpha + 1})\cdot \vec R} \ket{\psi^n},
\end{equation}
where $\vec R$ is the position operator, $\vec q_\alpha = \pi (\delta_{\alpha,1} + \delta_{\alpha,2},\delta_{\alpha,2} + \delta_{\alpha,3})$ for $\alpha = 0\ldots 3$, and where $\psi^m$ are the eigenfunctions of the system with periodic boundary conditions. By use of these matrices, the Chern number is then obtained through the equation
\begin{equation}
\mc C = \frac{1}{2\pi} \sum_m \arg(\lambda_m),
\end{equation}
with $\lambda_m$ being the complex eigenvalues of the matrix $C_{01}C_{12}C_{23}C_{30}$.

\section{Conductance}

A quantized thermal conductance is the topological response of the studied system. This observable also provides a powerful diagnostic tool in studying the robustness of the topological properties in the presence of low-energy excitations. Even in gapless systems, the conductance exhibits quantized values as long as the excitations below some finite energy are well-localized compared to the sample size. In the scattering theory framework, the thermal conductance can be expressed in terms of transmission of quasiparticles. The conductance of the system can be evaluated by connecting it to two semi-infinite leads, each modelled by a simple tight-binding Hamiltonian. The system is then treated as an $L_x\times L_y$ scattering area separating these leads. To make the numerical calculations easier, we fix a number of sites at two opposing edges and only connect these to the leads, while imposing no extra structure on the rest of the scattering area (see Fig.~\ref{app:fig:schematic}).

\begin{figure}
\includegraphics[width=0.8\linewidth]{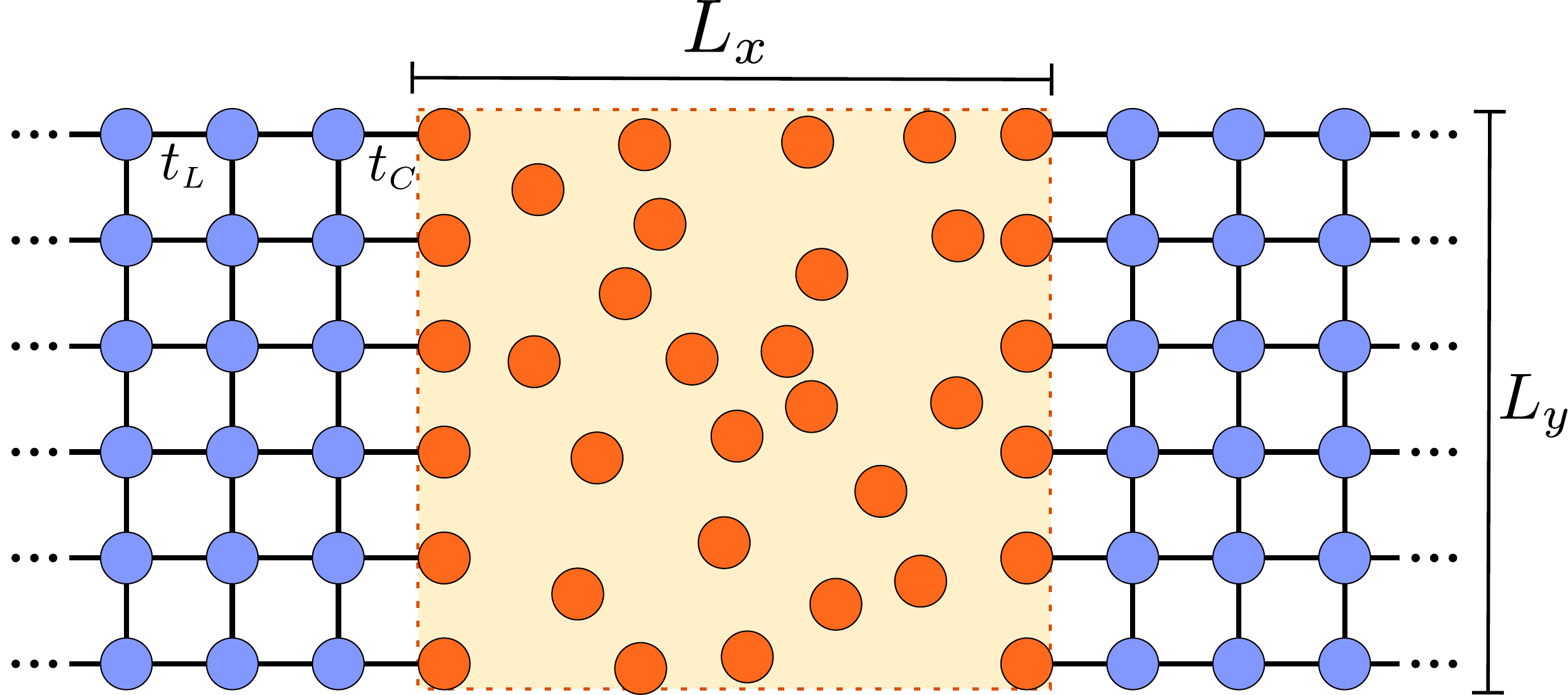}
\caption{Schematic figure of the conductance calculation setup. The box outlined in orange denotes the scattering area.} \label{app:fig:schematic}
\end{figure}

The retarded/advanced Green's function of the Shiba glass can be written as
\begin{equation}
G^{-1}_{r,a}(E) = E - H - \Sigma_{r,a},
\end{equation}
where $\Sigma$ is a self energy term originating from the coupling to the leads. This, in turn, can be found to be
\begin{equation}
\Sigma_r(E,m,n) = 
\begin{cases}
\frac{1}{L_y/a + 1} \left(\frac{t_C}{t_L}\right)^2 \sum_{k} \sin(k ma)\big(\varepsilon  - i\sqrt{4t_L^2-\varepsilon^2} \big) \sin(k na), & |\varepsilon| < 2t_L\\
\frac{1}{L_y/a + 1} \left(\frac{t_C}{t_L}\right)^2 \sum_{k} \sin(k ma)\big(\varepsilon  - \sgn \varepsilon\sqrt{\varepsilon^2 - 4t_L^2} \big) \sin(k na), & |\varepsilon| > 2t_L,
\end{cases}
\end{equation}
where $ma,na$ are the $y$ coordinates of sites connected to the leads, while $t_C$ and $t_L$ are the hopping parameter between the lead and the Shiba glass, and the hopping parameter within the lead, respectively. Furthermore, we have defined
\begin{equation}
\varepsilon = E - 2t_L\cos(ka).
\end{equation}
In the above expressions, $k$ takes values $k(j) = \frac{j\pi}{a(N_y+1)}$ for $j = 1\ldots N_y$, where $N_y$ denotes the number of fixed sites on each side of the scattering region. The thermal conductance quanta through the scattering area can then be calculated as
\begin{equation}
g = \TR\left[\Gamma_L G_r^{LR} \Gamma_R G_a^{RL} \right].
\end{equation}
Here, we have defined $\Gamma_{L,R} = -2\IM \Sigma_{L,R}$ as well as the matrices $G_{r,a}^{LR}$ and $G_{r,a}^{RL}$. The latter two are the subblock of the Green's function connecting the leftmost and rightmost edges of the sample -- in our setup, the matrix elements connecting the fixed sites on the opposite sides of the Shiba glass.

\section{Energy vs mobility gap of the Shiba glass in large systems}

\begin{figure}
\includegraphics[width=0.95\linewidth]{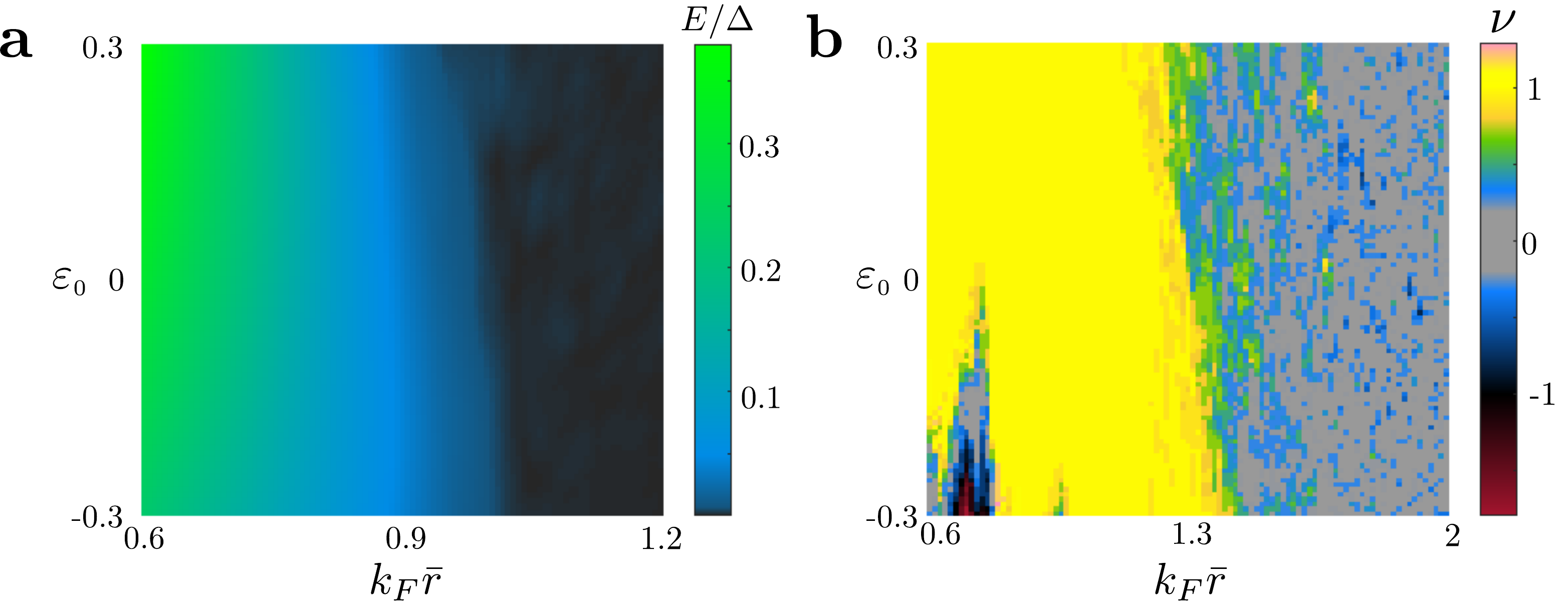}
\caption{\textbf{a,} Energy gap diagram for a finite system. In the topological phase, we observe a finite energy gap, while the trivial phase is on average gapless. The parameters used here are the same as in Fig.~2a in the main text. \textbf{b,} Chern number diagram for a Shiba glass with $k_F\xi = \frac{12}{5}\pi$ and $\lambda = 0.1$. Other parameters are the same as in Fig. 2a in the main text.}\label{Fig:EnergyGap}
\end{figure}
In the main text, we pointed out that in the infinite system limit, the Shiba glass is gapless but its topological character as manifested by quantized thermal conductance persists. The purpose of this section is to elucidate on that statement. Finite, reasonably small Shiba glass systems with high adatom density will with high probability be gapped in the bulk when in the topological phase, as can be seen in Fig.~\ref{Fig:EnergyGap}a. However, as the system size increases, the gap will scale down to zero. A system that is spatially infinite but has a finite density of randomly distributed adatoms will with probability 1 contain sparse regions, antipuddles, of any characteristic linear dimension $L$. For large enough $L$, this is enough to produce a local topologically trivial region even if the system on average is topological. The boundary of this topologically trivial region will then support some low-but-finite-energy excitations determined by the size of $L$ and vanishing as $L$ grows. Assuming that the moments are distributed independently with a fixed probability per unit area, the probability of getting locally trivial antipuddle configurations is exponentially suppressed as a function of antipuddle size $L$ and generally small for finite systems with densities that put them deep in the topological phase.  However, even for finite systems it is possible to create sizable such regions when placing sites randomly on a substrate.

To illustrate the effect of antipuddles, we have considered finite configurations which contain an empty region of specified $L$ and studied the properties of such systems. The results for one such configuration can be seen in Fig.~\ref{Fig:CustomHole}, where we have plotted its density of states (DOS) and thermal conductance. We have also compared both quantities to the case where the hole is filled.
\begin{figure}
\includegraphics[width=0.97\linewidth]{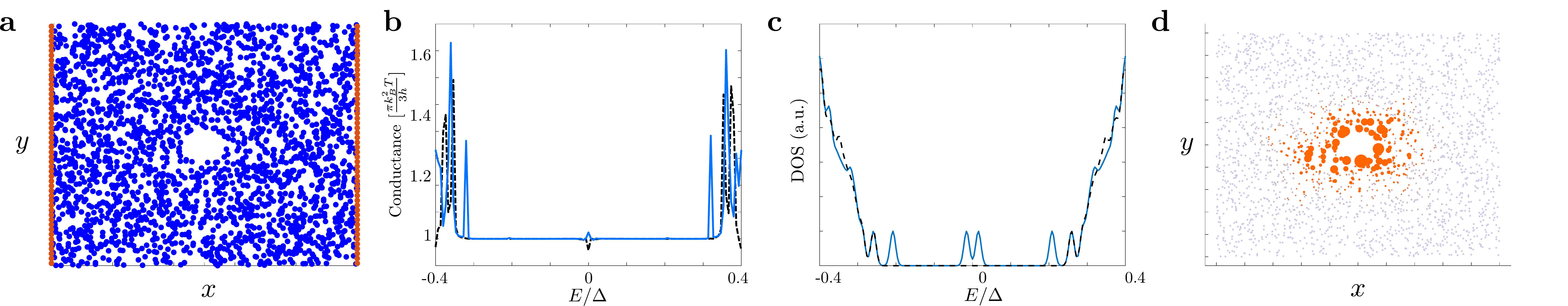}
\caption{Realization of a gapless system with nonzero mobility gap. \textbf{a,} A map of the lattice site configuration, with a small antipuddle in the middle. The sites colored in red are connected to semi-infinite leads. 
\textbf{b,} Heat conductance as a function of energy. From the figure, it can be observed that the heat conductance is quantized for energies $|E| \lesssim 0.3\Delta$. This quantized conductance plateau is only present in the topological phase. As a comparison, the dashed line corresponds to the conductance with the hole filled. As can be seen, the presence of the hole does not significantly impact the quantization.	
\textbf{c,} DOS for a system consisting of the disordered part of the scattering region, the blue sites in \textbf{a}, with periodic boundary conditions. The low energy states correspond to an edge mode around the antipuddle in the center. Periodic boundary conditions have been implemented to remove the effect of the heat conducting edge modes at the edges of the strip. From the figure it is clear that the system is gapless or nearly gapless, with DOS peaks at low energies. The dashed line corresponds to the DOS obtained for the same system with the hole filled; in this case, the low-energy peaks corresponding to antipuddle bound states are not present.
\textbf{d,} LDOS for the system described in \textbf{c}, i.e., the scattering region of \textbf{a} with periodic boundary conditions, integrated over energies $[-0.1\Delta,0.1\Delta]$. Observe that the low-energy states are bound at the antipuddle in the middle, and hence do not significantly disrupt the overall quantized heat conductance. }\label{Fig:CustomHole}
\end{figure}
The low-energy excitations, arising from the antipuddle in the center, are clearly visible in the enhanced DOS around $E=0$. However, the thermal conductance calculated for the configuration clearly exhibits a quantized plateau extending to $|E|=0.3\Delta$, explicitly showing that the low-energy excitations do not destroy the topological response of the system. This implies that the low-energy excitations are localized and the system possesses a well-defined mobility gap between $E=\pm0.3\Delta$ even though the energy gap is negligible. These results illustrate how it is possible that the energy gap scales to zero due to density fluctuations while the mobility gap, which is determined by the average density, will remain finite in the topological phase.  Thus, by increasing the system size but keeping the density fixed, we achieve a gapless system with a finite mobility gap and quantized thermal conductance.  As long as the density is sufficiently high that the system resides on the topological side of the phase boundary, the states around different antipuddles have negligible overlap and represent localized perturbations  that do not destroy the global topology of the system. 

\section{Phase diagrams at different parameters}

Metals generally have much clearer separation of Fermi wavelength and superconducting coherence length $k_F^{-1}\ll \xi $ compared to the proximity superconducting semiconductors. Also, metals typically have lower spin-orbit splitting at the Fermi level characterized by parameter $\lambda=\alpha_R/v_F$. To supplement the results presented in the main text, we consider how changing these two parameters affects the topological phase diagram. In general, decreasing $\lambda$ will have the effect of shifting the topological phase transition to higher densities, whereas increasing $\xi$ will have the opposite effect. These two effects compete and will, in effect, cancel each other out to some extent. In addition, increasing $\xi$ may introduce even new phase transitions, resulting in smaller regions within the overall topological parameter region that may have trivial topology or Chern numbers larger than unity. For sufficiently high densities these generally again vanish, leaving the system with a Chern number of magnitude one. An example of a phase diagram for higher $\xi$ and lower $\lambda$ can be seen in Fig.~\ref{Fig:EnergyGap}b; note that the system enters the topological phase at lower magnetic moment density (higher $\bar{r}$) than in the ones used in the main text, which is counteracted by the appearance of a new, smaller trivial parameter region within the topological region.

\end{widetext}
\end{document}